\documentclass[
    ,final            % use final for the camera ready runs
  ]
  {aipproc}

\layoutstyle{6x9}

\begin{document}

\title{Selecting the Optimal LHC Signatures for Distinguishing Models}


\author{Baris Altunkaynak}{
  address={Department of Physics, Northeastern University, Boston, MA 02115}
}

\begin{abstract}
An algorithm is developed which the goal of producing the most statistically significant signature list for distinguishing between two candidate models given a set of LHC observations.
\end{abstract}

\maketitle

Recently in \cite{Altunkaynak:2008ry} we investigated the LHC inverse problem first rigorously posed by Arkani-Hamed et al. in \cite{ArkaniHamed:2005px} and showed that non-collider data can be used to remove the degeneracies observed in the collider data. In \cite{Altunkaynak:2009tg} we attacked the same problem in the context of determining the non-universality in the gaugino sector by using LHC signatures. In this short note we summarize the statistical methods we utilized to optimize the set of signatures in order to minimize the integrated luminosity required to resolve the degeneracies.

We define a chi-square like distance function between any two models $A$ and $B$ as the metric of the signature space which is very similar to the one used in \cite{ArkaniHamed:2005px} as
\begin{equation} \label{DeltaS}
(\Delta S_{AB})^2 = \frac{1}{n} \, \sum_{i=1}^{n}  \left[ \frac{S_i^A - S_i^B}{\delta S_i^{AB}} \right]^2, 
\end{equation}
where $S_i$ is the $i^{th}$ counting signature and $\delta S_i^{AB}$ is the uncertainty of the numerator, i.e. the difference between the signatures which we will assume to contain only statistical errors. We can identify any signature $S_i$ with an ``effective'' cross section $\bar{\sigma}_i = S_i / L$ which includes the geometric
cuts that are performed on the data, the detector efficiencies, etc. At large integrated luminosity this converges to an ``exact'' cross section  $\sigma_i = \lim_{L \rightarrow \infty} \bar{\sigma}_i$. Rewriting the metric in terms of these effective cross sections gives us
\begin{equation} \label{DeltaS2}
(\Delta S_{AB})^2 = \frac{1}{n} \, \sum_{i=1}^n \left[
\frac{\bar{\sigma}_i^A - \bar{\sigma}_i^B}{\sqrt{ \bar{\sigma}_i^A /
L_A  + \bar{\sigma}_i^B / L_B }}\right]^2,
\end{equation}
where $L^A$ and $L^B$ are the integrated luminosities that are used to compute the effective cross sections.

We can obtain the statistical properties of this metric by replacing each signature (or effective cross section) by a random variable following a normal distribution. After this randomization, the effective cross sections simply become
\begin{equation} 
\bar{\sigma}_i = S_i^A / L_A = \sigma_i^A + \sqrt{\sigma_i^A
/ L_A} \, \, \, Z_A \, ,\label{sigrelate} 
\end{equation}
with a similar expression for the model $B$.

Substituting~(\ref{sigrelate}) into~(\ref{DeltaS2}) simply gives
\begin{eqnarray}
\label{ds2f} (\Delta S_{AB})^2 &=& \frac{1}{n}  \, \sum_{i=1}^n \frac{ \left[ \sigma_i^A -
\sigma_i^B + \sqrt{\frac{\sigma_i^A}{L_A} + \frac{\sigma_i^B}{L_B}}
Z_i \right]^2 }{ \frac{\sigma_i^A}{L_A} + \frac{\sigma_i^B}{L_B} + \sqrt{
\frac{1}{L_A^2} \frac{\sigma_i^A}{L^A}
+ \frac{1}{L_B^2} \frac{\sigma_i^B}{L^B} } Z_i' }  \nonumber \\
&\approx& \frac{1}{n}  \, \sum_{i=1}^n
\left[ \frac{\sigma_i^A - \sigma_i^B}{ \sqrt{\frac{\sigma_i^A}{L_A}
+ \frac{\sigma_i^B}{L_B}} } + Z_i'' \right]^2 \, , 
\end{eqnarray}
where $Z_i$, $Z_i'$ and $Z_i''$ are independent normally distributed random variables and assuming all $Z_i''$ are independent, i.e. our $n$ signatures are independent from each other, $(\Delta S_{AB})^2$ is itself a random variable having a non central chi-square distribution
\begin{equation}
P(\Delta S^2) = n \, \chi_{n,\lambda}^2(n \Delta S^2)\, ,
\label{prob}
\end{equation}
where $\lambda$ is the non-centrality parameter which is given by
\begin{equation}
\lambda = \sum_{i=1}^n \frac{(\sigma_i^A - \sigma_i^B)^2}{\sigma_i^A
/ L_A  + \sigma_i^B / L_B }\, . \label{lambdadef} 
\end{equation}
Here, $\lambda=0$ ($\neq 0$) corresponds to comparing a model to itself (to a different model) by using two sets of independent measurements. 
\begin{figure}[]
\includegraphics[scale=1,angle=0]{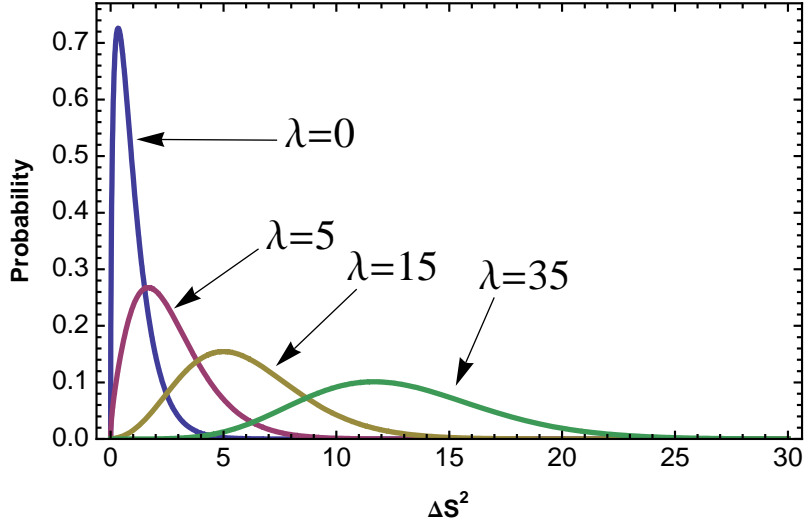}
\caption{Plot of distribution in $\Delta S^2$ values for $n=3$ and various $\lambda$ values.}
\label{distribution}
\end{figure}
Figure \ref{distribution} shows how the $(\Delta S_{AB})^2$ distribution favors larger values as $\lambda$ increases. Since our goal is to tell apart two models, we want the possible $(\Delta S_{AB})^2$ values we will get from this comparison to be safely away from the possible values we get by comparing a model to itself, i.e. $\lambda = 0$ case. If we quantify this safety condition as the requirement that $(100\times p)$\% of the distributions do not overlap, i.e. $(100\times p)$\% of the values we get by comparing the same model to itself are less than $(100\times p)$\% of the values we get by comparing two different models, we obtain the following equations
\begin{eqnarray}
p &=& \int_0^{\gamma} n \, \chi_{n,\lambda=0}^2(n \Delta
S^2)\, d(\Delta S^2) \quad \rightarrow \quad \Gamma \left(\frac{n}{2},\, \frac{n}{2} \, \gamma\right)
= \Gamma \left(\frac{n}{2}\right) (1-p)\, \\
p &=& \int_{\gamma}^{\infty} n \, \chi_{n,{\lambda_{\rm min}}}^2(n \Delta
S^2)\, d(\Delta S^2), \label{condition}
\end{eqnarray}
which can be solved numerically to compute a $\lambda_{\rm min}$ value (see Table \ref{lambdamin}) for every number of signatures $n$ and the non-overlap fraction (or confidence level) $p$. Here $\gamma$ is the $(\Delta S)^2$ cut-off value for which $(100\times p)$\% of the values we get by comparing a model to itself is less than this value and this condition gives us Eqn (7) which can be solved numerically to compute $\gamma$. Then this $\gamma$ value is used as the lower cut-off for the next equation which is solved again numerically to compute $\lambda_{\rm min}$.

The condition for two models to be distinguishable is simply $\lambda > \lambda_{\rm min}$. In this inequality $\lambda_{\rm min}$ is just a numerically computed number which is independent of the physics involved in the collider experiment and all the physics is in $\lambda$ which is a function of cross sections given by each signature.

\begin{table}[]
\centering
\begin{tabular}{c|c|c|c|c} 
 & \multicolumn{4}{c}{Confidence Level $p$}\\ \hline
 n & 0.95 & 0.975 & 0.99 & 0.999  \\
 \hline
 1 & 12.99 & 17.65 & 24.03 & 40.71 \\
 2 & 15.44 & 20.55 & 27.41 & 44.99 \\
 3 & 17.17 & 22.60 & 29.83 & 48.10 \\
 4 & 18.57 & 24.27 & 31.79 & 50.66 \\
 5 & 19.78 & 25.71 & 33.50 & 52.88 \\
 6 & 20.86 & 26.99 & 35.02 & 54.88 \\
 7 & 21.84 & 28.16 & 36.41 & 56.71 \\
 8 & 22.74 & 29.25 & 37.69 & 58.40 \\
 9 & 23.59 & 30.26 & 38.89 & 59.99 \\
 10 & 24.39 & 31.21 & 40.02 & 61.48 \\
\hline
\end{tabular}
\caption{List of
$\lambda_{\rm min}(n,p)$ values for various values of the parameters
$n$ and $p$.} \label{lambdamin}
\end{table}
Let us assume now that ``model $A$'' is the experimental
data, which corresponds to an integrated luminosity of $L^{\rm
exp}$, and ``model $B$'' is the simulation with integrated
luminosity $L^{\rm sim} = q L^{\rm exp}$. We might imagine that $q$
can be arbitrarily large, limited only by computational
resources. Let us make one final
notational definition
\begin{equation} 
R = \sum_{i=1}^N
\frac{ (\sigma_i^{\rm exp} - \sigma_i^{\rm sim})^2 }{ \sigma_i^{\rm exp} + \frac{1}{q} \,
\sigma_i^{\rm sim}} \, , \label{Rdef} 
\end{equation}
then we can compute the minimum amount of luminosity required for two models to be distinguishable which is given by
\begin{equation}
L_{\rm min} = \frac{\lambda_{\rm
min}(n,p)}{R}\, . \label{Lmin}
\end{equation}

If the two models we want to compare are very similar in all the channels (signatures) we consider, then $R$ will be small and $L_{\rm min}$ will be large. If on the other hand the models are very different $R$ will be large and $L_{\rm min}$ will be small. This is of course what we expect, i.e. similar models require more integrated luminosity to distinguish.

Now the question is how to make $L_{\rm min}$ as small as possible. We see from Table \ref{lambdamin} that $\lambda_{\rm min}$ increases as $n$ increases and since $R$ is a sum of positive quantities it increases with $n$ as well. Therefore using more signatures does not necessarily help in distinguishing models and, moreover, the signature space is not big enough (or at least the relevant part of the signature space, see \cite{ArkaniHamed:2005px}) to allow multiple independent directions. It is easy to see the orthogonality of signatures such as number of events with 1 lepton and 2 leptons, but for more general cases, such as kinematic histograms which we can integrate between limits that are also optimized to increase distinguishability, we need to compute the correlation coefficient between different signatures $a$ and $b$ which is given by
\begin{equation} 
\rho_{ab} = \frac{{\rm cov}(a,b)}{{\rm var}(a) {\rm var}(b)}
\approx  \frac{ \frac{1}{N} \sum_k \left[
\bar{\sigma}^k_a - \sigma_a \right] \left[\bar{\sigma}_b^k -\sigma_b
\right] } {\sqrt{ \frac{1}{N} \sum_k \left[\bar{\sigma}_a^k -
\sigma_a \right]^2} \sqrt{ \frac{1}{N} \sum_k \left[\bar{\sigma}_b^k
- \sigma_b \right]^2}}\, \quad \quad \textrm{for large }  N , \label{correlation} 
\end{equation}
where the $\bar{\sigma}^k$ represent the individual results obtained
from each of the $N$ cross section measurements, labeled by the
index $k$. This correlation matrix $\rho_{ab}$ then can be used to determine the compatible observables, i.e. the ones which are not correlated with each other with more than some fixed threshold $\epsilon$. This gives us the adjacency matrix of a graph which we define as
\begin{equation} C_{ab} = \left\{
\begin{array}{rl}
 1 & \textrm{if } |\rho_{ab}| \leq \epsilon \\
 0 & \textrm{if } |\rho_{ab}| > \epsilon \, .
\end{array} \right.
\label{Cmatrix} \end{equation}
Now finding the compatible observables is equivalent to finding all the complete subgraphs (or `clique') of that graph which is a well known problem in graph theory. All these complete subgraphs give us an $L_{\rm min}$ value and obviously the one giving the minimum of all these graphs contains the list of the signatures we want to combine together.

\begin{theacknowledgments}
The author is supported by National Science Foundation Grant PHY-0653587.
\end{theacknowledgments}

\end{document}